\documentclass[pdflatex,sn-mathphys-num]{sn-jnl}

\usepackage{graphicx}%
\usepackage{multirow}%
\usepackage{amsmath,amssymb,amsfonts}%
\usepackage{amsthm}%
\usepackage{mathrsfs}%
\usepackage[title]{appendix}%
\usepackage{xcolor}%
\usepackage{textcomp}%
\usepackage{manyfoot}%
\usepackage{booktabs}%
\usepackage{algorithm}%
\usepackage{algorithmicx}%
\usepackage{algpseudocode}%
\usepackage{listings}%
\usepackage{array}
\usepackage{ragged2e}
\usepackage{multirow}
\usepackage{colortbl}
\usepackage{tabularx}
\usepackage{tblr-extras}
\usepackage{makecell}
\usepackage{tabularray}
\usepackage{changepage}
\usepackage{nameref,hyperref}
\usepackage[noabbrev,nameinlink,capitalise]{cleveref}
\usepackage{array}
\usepackage{url}

\DefTblrTemplate{caption-tag}{default}{\textbf{Table\hspace{0.25em}\thetable}}
\DefTblrTemplate{caption-sep}{default}{\enskip}

\newlength\savedwidth

\definecolor{Gray}{rgb}{0.501,0.501,0.501}

\usepackage[aboveskip=1pt,labelfont=bf,labelsep=period,justification=raggedright,singlelinecheck=off]{caption}
\renewcommand{\figurename}{Figure}

\begin{document}

\title[Article Title]{Journalists are most likely to receive abuse: Analysing online abuse of UK public figures across sport, politics, and journalism on Twitter}

\author*[1]{\fnm{Liam} \sur{Burke-Moore}}\email{lburke-moore@turing.ac.uk}
\equalcont{These authors contributed equally to this work.}

\author[1]{\fnm{Angus} \sur{R. Williams}}\email{arwilliams@turing.ac.uk}
\equalcont{These authors contributed equally to this work.}

\author[1]{\fnm{Jonathan} \sur{Bright}}\email{jbright@turing.ac.uk}

\affil[1]{\orgdiv{Public Policy}, \orgname{The Alan Turing Institute}, \orgaddress{\city{London}, \postcode{NW1 2DB}, \country{United Kingdom}}}

\abstract{Engaging with online social media platforms is an important part of life as a public figure in modern society, enabling connection with broad audiences and providing a platform for spreading ideas. However, public figures are often disproportionate recipients of hate and abuse on these platforms, degrading public discourse. While significant research on abuse received by groups such as politicians and journalists exists, little has been done to understand the differences in the dynamics of abuse across different groups of public figures, systematically and at scale. To address this, we present analysis of a novel dataset of 45.5M tweets targeted at 4,602 UK public figures across 3 domains (members of parliament, footballers, journalists), labelled using fine-tuned transformer-based language models. We find that MPs receive more abuse in absolute terms, but that journalists are most likely to receive abuse after controlling for other factors. We show that abuse is unevenly distributed in all groups, with a small number of individuals receiving the majority of abuse, and that for some groups, abuse is more temporally uneven, being driven by specific events, particularly for footballers. We also find that a more prominent online presence and being male are indicative of higher levels of abuse across all 3 domains.}


\maketitle

\section{Introduction}\label{section:intro}
It is by now commonplace for public figures (well known individuals such as celebrities, sports stars, journalists, politicians) to maintain an active presence on social media platforms such as Instagram, Facebook and X (Twitter). Such platforms allow them to build up a personal `brand' and connection with an audience through the generation of a kind of informal, parasocial intimacy \cite{dai_vicariously_2018}; a brand that can then be leveraged for a wide variety of different professional purposes, from securing sponsorship deals \cite{balliauw_identifying_2020} to influencing opinions \cite{weeks_online_2017} or distributing ideas \cite{bossio_mapping_2016}. Many researchers have highlighted the potential positive consequences of the fact that access to public status is now available through social media channels, in particular by allowing voices into the public sphere who previously would have remained marginalised \cite{tucker_liberation_2017}.

However, whilst these positive consequences are important, much of the research on high profile figures on social media has also focused on a negative aspect: the frequency with which these figures receive hateful and abusive messages, facilitated through the peer-to-peer nature of social media. Such messages can be personally distressing for the individuals involved \cite{deavours_reciprocal_2023, thompson_digital_2022}, and may lead them to limiting their online presence in order to avoid receiving such messages \cite{cover_protecting_2024}, which in turn will degrade the quality of public discourse and limit the ability for the public to engage with, for example, elected officials. High levels of abuse can also cause more widespread consequences for those witnessing the abuse, who may conclude that public debate is a hostile, angry environment that they should stay away from \cite{stevens_women_2024}. 

Though research on abuse of public figures is widespread, it is also largely siloed, with individual efforts looking at (for example) abuse levels towards politicians, or journalists, or a certain type of celebrity. As methodology can differ across these individual studies, direct comparisons can be complex. We therefore know little about the extent to which dynamics of abuse are similar across different domains, and across demographics within domains. We attempt to advance the debate about abuse towards online public figures by providing a measurement of the extent of abuse faced by three key United Kingdom (UK) based groups (members of parliament, journalists and footballers), and a comparison of the dynamics of abuse between them.

In this paper, we present a cross-group analysis of a novel dataset of $45.5$M tweets targeted at $4,602$ UK public figures across 3 domains, collected between 2021 and 2023, applying our previous work fine-tuning pre-trained transformer models to classify abusive tweets. We find that MPs as a group receive more abuse than footballers or journalists, but show through statistical modelling that abuse may be more of an intrinsic feature of being a journalist than other domains. We also find that a more prominent online presence and being male are factors predictive of higher levels of abuse, and that abuse is unevenly distributed both individually and temporally across all groups. 

\section{Related Work}\label{section:relatedwork}
This paper focuses on the issue of abuse towards public figures on social media. We define abuse as content that threatens, insults, derogates, mocks or belittles an individual or their identity \cite{williams_dodo_2024}. This is a broad reaching definition that includes but is not limited to more severe forms of abuse that may constitute `hate speech' (hate focused on protected characteristics \cite{EqualityAct2010}), also accounting for generic toxicity.

Whilst abuse has been a constant feature of public life, a wide variety of research has found that it appears to be especially prevalent in online discussions. Some of the earliest work on the internet remarked on the apparent prevalence of `flaming' \cite{lea_flaming_1992}, with research continuing to this day about aggressive and uncivil online comments in forums and discussion sections of websites \cite{kumar_toxic_2023, masullo_toxic_2023}. A considerable further body of research has tried to explain why online environments seem to be so much more hostile than offline ones \cite{bor_online_2019}, highlighting factors such as anonymity \cite{zimmerman_online_2012}, reduced empathy, and group dynamics where witnessing (and receiving) abuse makes one more likely to create it \cite{wachs_associations_2019}. 

Early research on online abuse regarded it largely as an interpersonal phenomenon, exchanged amongst nascent internet communities that were at the time a niche pursuit \cite{spitzberg_cyberstalking_2002}. However as the internet itself grew into the major means of online societal communication, a further focus developed in terms of the fact that `public figures' also started to become targets. In this paper, we use the term `public figures' to define those whose profession compels them to seek recognition amongst the public and who therefore become known to a potentially wide section of the community. They communicate with an audience of individuals who are unknown to them. Public figures include celebrities, sports stars and famous politicians who might be known to an audience of millions. However they may also include local journalists or members of parliament who might have much lower name recognition but nevertheless have a public face. As Xu et al. \cite{xu_parasocial} show, public figures now exist on all scales, from `traditional' celebrities known to millions to `micro-celebrities'. Such figures enter into `parasocial' relationships with members of the wider public: a type of relationship which feels intimate and personal on some level to audience members despite the fact that the public figure themselves is unlikely to have met all of (or even a small fraction of) the audience members with whom they have such a relationship \cite{horton_mass_1956}. 

While public figures have always been a fact of social life, the way they work has changed dramatically with the rise of the internet, and especially social media. These platforms, which provide the possibility of such figures communicating in a relatively direct way with their audiences (circumventing to an extent the filtering mechanisms of the press), have revolutionised what it means to be a public figure, giving rise to a wide variety of new ways of forming parasocial relationships \cite{bond_following_2016, lim_role_2020}. A presence on at least some social media platforms is by now arguably a requirement of many public facing professions (such as politics and journalism), or at least a highly important way of advancing professional life or monetising status through (for example) endorsements \cite{aw_celebrity_2020}.

While access to these platforms arguably represents a boon in many senses, the levels of abuse public figures receive on them is also a source of increasing concern. In this study, we address three different categories of public figure in particular. We look at professional sports stars (in particular football players), journalists, and politicians. This selection does not, of course, address all potential types of public figures (for example, musicians and actors are obvious absences from the list). However, it does offer important variety, with three very different professions with different audiences to communicate with, all united only by the fact that they engage with people across public facing social media. Each of these categories has attracted a considerable amount of research on levels of abuse, which we will review in turn below. What is missing, which we provide here, are studies that address multiple categories in the same framework, and thus provide a more general view on the dynamics of abuse. 

\paragraph{Football}

Abuse towards professional athletes (and other professionals such as referees) has long been part of the sporting industry \cite{burdsey_race_2011}, and in the past has been associated with multiple campaigns launched by the industry itself to attempt to stamp it out \cite{webb_distribution_2017}. While there was some perceptions that these campaigns had been partially successful, the rise of social media as a forum for the self-presentation of footballers for many brought about a kind of regression with abuse once again rife, also closely associated with the issue of racism \cite{kilvington_tackling_2019}. Empirical work has consistently documented relatively high levels of abuse towards footballers \cite{vidgen_tracking_2022, grez_new_2022}. However the vast majority of quantitative studies have, to our knowledge, been directed towards male sports stars, despite obvious press attention to abuse towards female athletes as well \cite{staff_karen_2021}. Concerns about abuse (especially racist abuse) being directed towards footballers are based of course on the mental health and wellbeing of the players themselves, with players having even contemplated suicide when being on the receiving end of it \cite{staff_karen_2021} and family members also feeling an incredible amount of strain \cite{noauthor_rio_nodate}. Furthermore, due to the highly mediatized nature of the phenomenon, there are concerns that witnessing of online abuse may serve to normalise it in wider society. 

\paragraph{Politics}

Volumes of abuse towards professional politicians have also been a subject of considerable research interest. The world of professional politics is of course a combative one, with threat and harassment unfortunately a part of life for professional politicians of all types \cite{bjorgo_patterns_2022, marijnissen_coping_2020}. The online arena seems to be an extension of this trend, with a wide variety of work documenting the high levels of abuse and vitriol directed towards elected officials \cite{ward_turds_2020, gorrell_online_2020}, and some also arguing that the problem is increasing over time \cite{gorrell_twits_2018, gorrell_mp_2020}. 

One of the key debates in this area is whether male and female politicians experience different volumes and types of abuse, with some studies not identifying gender differences \cite{ward_turds_2020, gorrell_which_2020}, whilst others have problematised this type of finding \cite{harmer_digital_2021, collignon_increasing_2021, phillips_gender_2023, erikson_three_2021, fuchs_normalizing_2020} or shown mixed results \cite{van_der_vegt_gender_2024}. The recent resignations of high profile female politicians, citing patterns of abuse received, seem highly significant in this regard \cite{noauthor_jacinda_nodate, wagner_tolerating_2022}. A similar but smaller body of literature has also sought to highlight religious and racial differences \cite{gorrell_race_2019}. Another key debate is the reasons for abuse \cite{gorrell_which_2020}, with some arguing that periodic news attention to different topics is a key driver and others pointing to differences in the profile of the individuals in question \cite{gorrell_twits_2018}. These issues are critical not only in terms of effects on politicians themselves and wider public discourse, but also in terms of concerns about impacting on the representativeness of democracy as a whole. 

\paragraph{Journalism}

The impact of online abuse on the journalistic profession has also attracted a considerable amount of scrutiny (including noting the clear crossovers with the previous two domains in terms of journalists covering both sports and politics). Findings are in a sense similar to the other two domains: first, a wide variety of scholarship has claimed the problem is serious and widespread \cite{binns_fair_2017, relly_online_2021, waisbord_mob_2020}, as well as being connected to real world acts of violence, a subject of particular concern as many journalists lack the security protections provided to politicians (though this is not to say that politicians do not also frequently experience violent attacks). Diverging patterns of abuse between men and women have also been a frequent area of study \cite{simoes_online_2021, unesco_online_2020, posetti_7_nodate}: though unlike in the political arena, greater levels of abuse directed towards women has been a clear and consistent finding \cite{lewis_online_2020, binns_fair_2017}.

Personal visibility (as opposed to newsroom visibility) is suggested as another factor of the abuse of journalists \cite{lewis_online_2020}, and many studies have also linked it to broad societal factors such as the rise of populism \cite{waisbord_mob_2020, relly_online_2021}. Temporal factors have also been considered, with work describing online abuse towards journalists as both a chronic problem and one that is also likely to be boosted by individual events \cite{holton_not_2023}. Some of the feared consequences of abuse for journalists are also somewhat similar to the political domain: that the distress and psychological burden created by abuse patterns will drive people out of the public domain \cite{ferrier_trollbusters_2018, binns_fair_2017}. However, authors have also noted that this abuse may create a more general perception in the eyes of journalists themselves that news audiences are irrational and low quality \cite{lewis_online_2020}.  \bigskip

One of the most significant and yet under-explored things emerging from all of this work is that, despite the great differences in the profession and style of work these different public figures are employed in, many similar patterns and claims about online abuse have emerged. However, what the field as yet lacks is comparative work looking at different professions to tie these observations together (one notable exception is \cite{delisle_large-scale_2019}, though this looks only at female journalists and politicians). We hence lack knowledge about what features of abuse are unique to a given professional context and what are more general features of online public life as a whole. 

In this article, we seek to remedy this deficit, by measuring levels of abuse across our three different domains of interest. We structure our enquiry in terms of three key questions: \smallskip

\noindent\textbf{Distribution of abuse:} do all domains experience similar levels of abuse, and is this abuse distributed amongst people within the domain in similar ways? \smallskip

\noindent\textbf{Temporal patterns:} is abuse a generally stable features of domains, or does it fluctuate and respond to events? \smallskip

\noindent\textbf{Factors linked to abuse:} how does abuse vary with the activity of public figures? After accounting for other potential factors linked to abuse, to what extent is abuse an intrinsic feature of a domain? \smallskip

Our aim is to describe the dynamics of online hate and abuse towards public figures in a way that is not entirely dependent on the domain or field of study.

\section{Methods}\label{section:methods}
\subsection{Platform Selection}
\label{section:platform}
In this paper we make use of data collected from the social media platform Twitter/ X (we refer to ``Twitter'' exclusively given data collection took place primarily before the rebrand to ``X'' ). As a platform focused on broadcasting messages, with no requirement for reciprocal following before messages are exchanged, Twitter has long been a forum where public figures have maintained an active presence and broadcast messages to an audience. It hence represents an ideal choice of venue for our study. It is worth noting that other platforms are also being used by public figures, with the footballers in our study also highly present on Instagram, for example. However, there is no other platform that is widely used by all the groups in our study.  

\subsection{Target Group Selection}
\label{section:group}

We select 3 domains of public figures to study: professional football players, members of parliament, and journalists. We delineate public figures into male and female groups (this study is limited to binary gender, given the low prevalence of individuals identifying as non-binary or other genders within these groups). This gives us 6 individual groups across 3 domains and 2 genders.

We source lists of public figures from official sources where available, and filter lists down to include only those with an official Twitter account. Full details are visible in \cref{appendix:datacollection}. The final, total number of individual public figures, present on Twitter, across all domains and demographics, included in this study are visible in \cref{table:pfaccounts}. Immediately apparent is the minority of female public figures across all domains. This is mirrored by follower accounts, where the average man individual public figure has a higher follower count than the average woman, and the majority of the top 10 most followed individuals in each domain-demographic pair are men. This could be seen as a feature of the domains themselves (and society in general):  while the popularity of women's football grows, men's football receives more widespread engagement \cite{FIFA2019}, the the balance of female MPs was below 20\% until 2006 \cite{parliament_women_2022}, and journalism has been shown to be an industry dominated by men \cite{eddy_women_2023}. 

\begin{table}
% \footnotesize
\centering
\begin{talltblr}[caption={Counts of public figure Twitter accounts.}, label={table:pfaccounts}]{
  width = \linewidth,
  cell{1}{1} = {r=2}{},
  cell{1}{2} = {c=2}{c},
  cell{1}{4} = {r=2}{},
  cell{3}{2} = {r},
  cell{3}{3} = {r},
  cell{3}{4} = {r},
  cell{4}{2} = {r},
  cell{4}{3} = {r},
  cell{4}{4} = {r},
  cell{5}{2} = {r},
  cell{5}{3} = {r},
  cell{5}{4} = {r},
  % vline{2} = {1-2}{},
  % vline{4} = {1-2}{},
  % vline{2,4} = {3-5}{},
  hline{1,3,6} = {-}{},
}
\textbf{Domain}       & \textbf{Gender } &               & \textbf{Total} \\
                      & \textit{Female}  & \textit{Male} &                \\
Footballers           & 204              & 807           & 1,011          \\
Members of Parliament & 207              & 384           & 591            \\
Journalists           & 1,210            & 1,790         & 3,000          
\end{talltblr}
\end{table}

\subsection{Data Collection}
\label{section:collection}

Central to Twitter activity are primarily text-based posts called ``tweets''. Tweets can be replied to, creating chains or ``threads'', or can be reposted as ``Quote tweets'' or ``Retweets'' of other tweets. We are interested in the most direct form of communication targeted at public figures, and as such we only consider what we term \textit{``audience contact''} (AC) tweets: direct replies to a tweet from a public figure account, or top-level tweets (that aren't replies to other tweets) containing a mention of a public figure account. We present aggregate statistics based on these tweets and the labels assigned.

We use the Twitter API Filtered Stream endpoint and Full Archive Search endpoint (provided by the Twitter Academic API, no longer available) to collect all tweets that either contain a mention of a public figure account (including direct and indirect replies) or are quote tweets or retweets of tweets created by a public figure account. Data collection endpoint usage and time windows differed across domains and demographics, as outlined in \cref{table:windows}, due to the staggered nature of data collection for this project. All data collection ended on the 14th of March 2023, when API access was suspended. We filter the tweets collected to retain only tweets matching the \textit{audience contact} conditions, that are written in English, and contain text content aside from mentions and URLs. On collection, we extracted lists of public figure accounts mentioned within the tweet text, and created a clean version of the tweet text, replacing mentions of users with domain-specific tokens, and URLs with a URL token. The remaining ``valid'' \textit{audience contact} tweets (visible in \cref{table:windows}) for each domain-demographic pair are used for the modelling and analysis presented in this paper. 

\begin{table}[!h]
% \begin{adjustwidth}{-2.25in}{0in}
\centering
\begin{talltblr}[caption={Table of data collection windows and Tweet counts per domain and demographic groups.}, label={table:windows}]{
  colspec={X[.86,l]X[.94,l]X[.83,l]X[.82,l]X[.54,r]X[1.5,r]},
  cell{2}{1} = {r=2}{},
  cell{4}{1} = {r=2}{},
  cell{6}{1} = {r=2}{},
  hline{1-2,8} = {-}{},
  hline{4,6} = {-}{Gray},
}
\textbf{Domain} & \textbf{Demographic} & \textbf{Start Date} & \textbf{End Date} & \textbf{Total Days} & \textbf{Audience Contact Tweets}                \\
Footballers     & Men                  & 12/08/2021          & 14/03/2023        & 579                 & 7,398,876  \\
                & Women                & 12/08/2021          & 14/03/2023        & 579                 & 303,403 \\
MPs             & Men                  & 13/01/2022          & 14/03/2023        & 425                 & 18,741,751   \\
                & Women                & 13/01/2022          & 14/03/2023        & 425                 & 9,404,846  \\
Journalists     & Men                  & 01/08/2022          & 31/01/2023        & 183                 & 7,300,005  \\
                & Women                & 01/08/2022          & 31/01/2023        & 183                 & 2,343,299  
\end{talltblr}
% \end{adjustwidth}
\end{table}

We additionally collect all tweets authored by public figure accounts, again using the Twitter API Filtered Stream endpoint and Full Archive Search endpoint, in order to enable analysis around the activity of public figures and the relationship with abuse. We retain all of these tweets, regardless of language and content, and we do not label these tweets as abusive / not abusive.

\subsection{Abuse Classification}
\label{section:labelling}

We fine-tune pre-trained transformer-based language models for binary abuse classification for each public figure target group, using the same annotated data and annotation processes outlined in Vidgen et al.\cite{vidgen_tracking_2022} and Williams et al.\cite{williams_dodo_2024}.

All tweets are annotated with one of four labels: ``abusive'', ``critical'', ``neutral'', or ``positive''. Definitions of each class, and guidelines for annotators, are visible in \cref{fig:annotinstructions}. Here we define abuse as broad-reaching, including but not limited to hate speech, pertaining to any content that threatens, insults, derogates, mocks or belittles an individual or their identity \cite{williams_dodo_2024}. We collapse multi-class labels to binary labels (abuse / not abuse) for the models in this study. At least 7,000 tweets are annotated for each group: $1,000$ for the validation split, $3,000$ for the test split, and $3,000$ for the training split (more in the case of male footballers).

Initial rounds of annotation (the male and female footballers datasets) were done by crowdworkers, but, due to high levels of disagreement between crowdworkers, and therefore more expert annotation required, a small group of high-quality annotators was used to label the remaining datasets (MPs, Journalists).

For \textbf{male footballers}, we use a version of deBERTa-v3 \cite{he_deberta_2021} fine-tuned on $9,500$ tweets targeted at male footballers. This model is trained using an active learning process, starting with a sample of $3,000$ tweets, and using diversity and uncertainty sampling to select $2,000$ additional training entries to annotate over 3 rounds, plus one round of $500$ adversarial entries, as outlined in Vidgen et al.\cite{vidgen_tracking_2022}. Tweets were annotated by 3,375 crowdworkers. This model outperformed (F1 score on the male footballers test split) a model trained on the base $3,000$ male footballers training dataset, and an ensemble of two models trained on male footballers and female footballers data.

For \textbf{female footballers}, we use an ensemble of two fine-tuned versions of deBERTa-v3 \cite{he_deberta_2021}, one on tweets targeted at male footballers, the other on tweets targeted at female footballers. Both models were fine-tuned on $3,000$ tweets, as outlined in Williams et al.\cite{williams_dodo_2024}. Tweets were annotated by 3,513 crowdworkers. Output probabilities from the two models are averaged during inference to make classifications. This ensemble outperformed (F1 score on the female footballers test split) the model trained on the base $3,000$ female footballers training dataset.

For \textbf{MPs}, we use an ensemble of two fine-tuned versions of deBERTa-v3 \cite{he_deberta_2021}, one on tweets targeted at male MPs, the other on tweets targeted at female MPs. Both models were fine-tuned on $3,000$ tweets, as outlined in Williams et al.\cite{williams_dodo_2024}. Tweets were annotated by 23 high quality annotators. Output probabilities from the two models are averaged during inference to make classifications. This ensemble outperformed models trained on the base training datasets for both male and female MPs.

For \textbf{journalists}, we use a version of deBERTa-v3 \cite{he_deberta_2021} fine-tuned on $3,000$ tweets targeted at journalists, following the same processes outlined in Williams et al.\cite{williams_dodo_2024}. Tweets were annotated by 23 high quality annotators.

Model evaluation results are visible in \cref{table:modeleval}.

\begin{table}
\centering
\begin{adjustwidth}{-0.4in}{0in}

\begin{talltblr}[caption={Evaluation of models used for abuse classification.}, label={table:modeleval}]{
    rows = {
        valign=m,
    },
  width = 1.1\linewidth,
  colspec = {Q[75]Q[53]Q[240]Q[78]Q[67]Q[53]Q[50]},
  column{4} = {c},
  column{5} = {c},
  column{6} = {c},
  column{7} = {c},
  cell{2}{1} = {r=2}{},
  cell{4}{1} = {r=2}{},
  cell{4}{3} = {r=2}{},
  cell{4}{4} = {r=2}{},
  cell{6}{1} = {r=2}{},
  cell{6}{3} = {r=2}{},
  cell{6}{4} = {r=2}{},
  hline{1-2,8} = {-}{},
  hline{3} = {2-7}{Gray},
  hline{4,6} = {-}{Gray},
  hline{5,7} = {2,5-7}{Gray},
}
\textbf{Domain} & \textbf{Gender} & \textbf{Model Description}                                                                                                                     & \textbf{Train Size} & \textbf{F1 Score} & \textbf{Precision} & \textbf{Recall} \\
Footballers     & Men                  & deBERTa-v3 model fine-tuned on male footballer data through active learning \cite{vidgen_tracking_2022}                                        & 9,500              & 0.65              & 0.65               & 0.64            \\
                & Women                & Ensemble of 2 deBERTa-v3 models, one fine-tuned on male footballer data, the other on female footballer data \cite{williams_dodo_2024} & 2*3,000         & 0.62              & 0.79               & 0.51            \\
MPs             & Men                  & Ensemble of 2 deBERTa-v3 models, one fine-tuned on male MP data, the other on female MP data \cite{williams_dodo_2024}                 & 2*3,000         & 0.68              & 0.71               & 0.65            \\
                & Women                &                                                                                                                                                &                     & 0.61              & 0.67               & 0.56            \\
Journalists     & Men                  & deBERTa-v3 model fine-tuned on journalist data                                                                                                 & 3,000               & 0.62              & 0.61               & 0.64            \\
                & Women                &                                                                                                                                                &                     & 0.66               & 0.66                & 0.66             
\end{talltblr}
\end{adjustwidth}
\end{table}

\section{Results}\label{section:results}
We structure our analysis in terms of our three research questions.

\subsection{Distribution of abuse}
\label{section:dist_of_abuse}

We begin our analysis by assessing whether all domains experience similar levels of abuse, and look at whether this abuse is distributed amongst individual public figures within a domain in similar ways.

We present total tweet counts in \cref{table:total_metrics} and weekly average tweet counts in \cref{table:wkavg_metrics} (total counts are presented for completeness, weekly averages are used for analysis due to variable time windows between domains). We see that MPs have the highest weekly average rate of abuse, with 11.2\% of tweets received by male MPs being classified as abusive, and 9.1\% for female MPs. We also see that male footballers receive a higher average proportion of abuse containing identity-based slurs than any other group at 11\%. 32 journalists received no Tweets during the data collection window.

\begin{figure}[!h]
    \centering
    \begin{adjustwidth}{-0.2in}{0in}
    \includegraphics[width=1.05\linewidth]{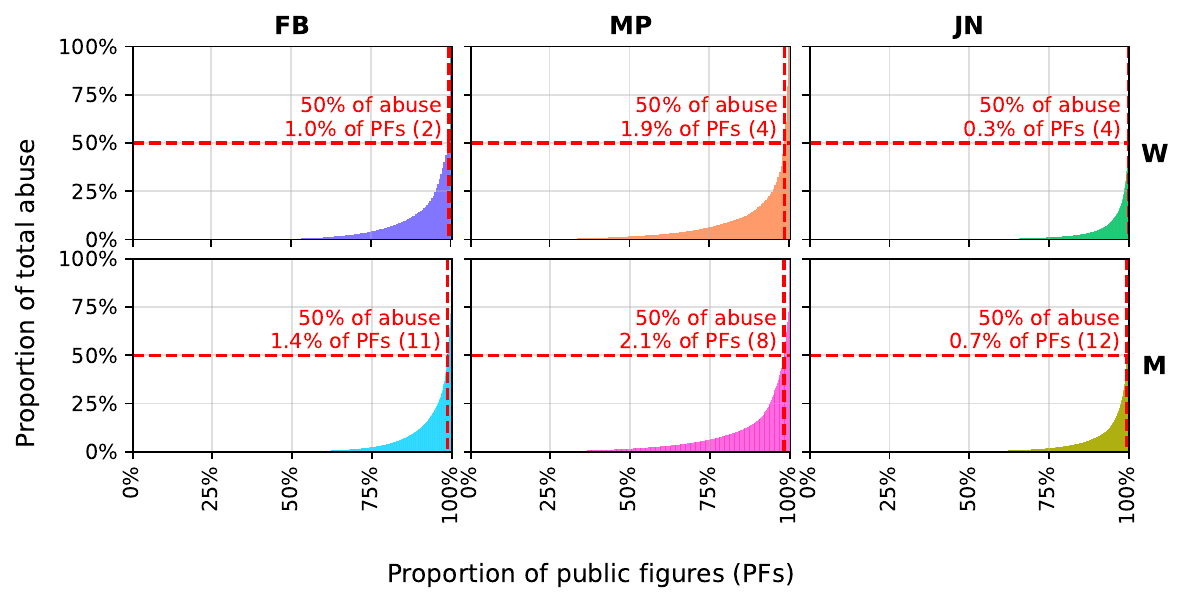}
    \caption{
    {Cumulative distributions of abuse by individual public figure.} \\ \small One column per domain, one row per gender. Proportion of the required top $N\%$ of public figures to reach 50\% of all abuse is annotated.} \label{fig:userdist}
    \end{adjustwidth}
\end{figure}

We present cumulative distributions of total abuse counts by individual public figures in \cref{fig:userdist}. We see that, across all domains and demographics, a small number of individuals receive a large proportion of the total abuse. For example, 50\% of abuse targeted at male MPs is directed towards just 2.1\% of all of those individuals. This observation holds for other domains and demographics, although differences can be partly explained by the differing number of public figures in each group.

\begin{table}[!t]
\centering
\begin{adjustwidth}{-0.9in}{0in}
\begin{talltblr}[caption={Total metrics per domains and demographic group}, label={table:total_metrics}]{
  width = 1.1\linewidth,
  colspec = {Q[l,55]Q[l,55]Q[r,65]Q[r,75]Q[r,82]Q[r,60]Q[r,80]},
  cell{2}{1} = {r=2}{},
  cell{2}{6} = {r},
  cell{2}{7} = {r},
  cell{3}{6} = {r},
  cell{3}{7} = {r},
  cell{4}{1} = {r=2}{},
  cell{4}{6} = {r},
  cell{4}{7} = {r},
  cell{5}{6} = {r},
  cell{5}{7} = {r},
  cell{6}{1} = {r=2}{},
  cell{6}{6} = {r},
  cell{6}{7} = {r},
  cell{7}{6} = {r},
  cell{7}{7} = {r},
  hline{1-2,8} = {-}{},
  hline{4,6} = {-}{Gray},
  row{1} = {font=\bfseries},  % Bold for the header row
}
Domain          & Demographic & Tweets Authored & Total Tweets Received & Abusive Tweets Received & Abuse Proportion & Identity-Abuse Proportion \\
Footballers     & Men         & 31,776                   & 7,398,876                 & 186,085                    & 2.5\%             & 11.3\%                    \\
                & Women       & 13,405                   & 303,403                  & 2,599                       & 0.9\%             & 5.8\%                     \\
MPs             & Men         & 73,535                & 18,741,751                & 2,131,022                   & 11.4\%            & 5.9\%                     \\
                & Women       & 46,190                   & 9,404,846                & 884,493                   & 9.4\%             & 4.9\%                     \\
Journalists     & Men         & 464,132                & 7,300,005                & 535,631                   & 7.3\%             & 5.9\%                     \\
                & Women       & 149,879                 & 2,343,299                 & 159,516                    & 6.8\%             & 5.0\%                     \\
\end{talltblr}
\newline
\vspace*{1em}
\newline
\begin{talltblr}[caption={Weekly average metrics per domain and demographic group.}, label={table:wkavg_metrics}]{
  width = 1.1\linewidth,
  colspec = {Q[l,55]Q[l,55]Q[r,65]Q[r,75]Q[r,82]Q[r,60]Q[r,80]},
  cell{2}{1} = {r=2}{},
  cell{2}{6} = {r},
  cell{2}{7} = {r},
  cell{3}{6} = {r},
  cell{3}{7} = {r},
  cell{4}{1} = {r=2}{},
  cell{4}{6} = {r},
  cell{4}{7} = {r},
  cell{5}{6} = {r},
  cell{5}{7} = {r},
  cell{6}{1} = {r=2}{},
  cell{6}{6} = {r},
  cell{6}{7} = {r},
  cell{7}{6} = {r},
  cell{7}{7} = {r},
  hline{1-2,8} = {-}{},
  hline{4,6} = {-}{Gray},
  row{1} = {font=\bfseries},  % Bold for the header row
}
Domain          & Demographic & Tweets Authored & Total Tweets Received & Abusive Tweets Received & Abuse Proportion & Identity-Abuse Proportion \\
Footballers     & Men         & 942             & 88,082                & 2,215                   & 2.5\%             & 11.0\%                    \\
                & Women       & 201             & 3,528                 & 30                      & 0.6\%             & 5.6\%                     \\
MPs             & Men         & 1,559           & 302,286               & 34,371                  & 11.2\%            & 6.0\%                     \\
                & Women       & 931             & 151,691               & 14,266                  & 9.1\%             & 5.0\%                     \\
Journalists     & Men         & 18,360          & 270,371               & 19,838                  & 7.2\%             & 5.9\%                     \\
                & Women       & 6,034           & 86,789                & 5,908                   & 6.8\%             & 5.0\%                     \\
\end{talltblr}
\end{adjustwidth}
\end{table}

\begin{figure}[!h]
    % \begin{adjustwidth}{-0.6in}{0in}
    \includegraphics[width=1\linewidth]{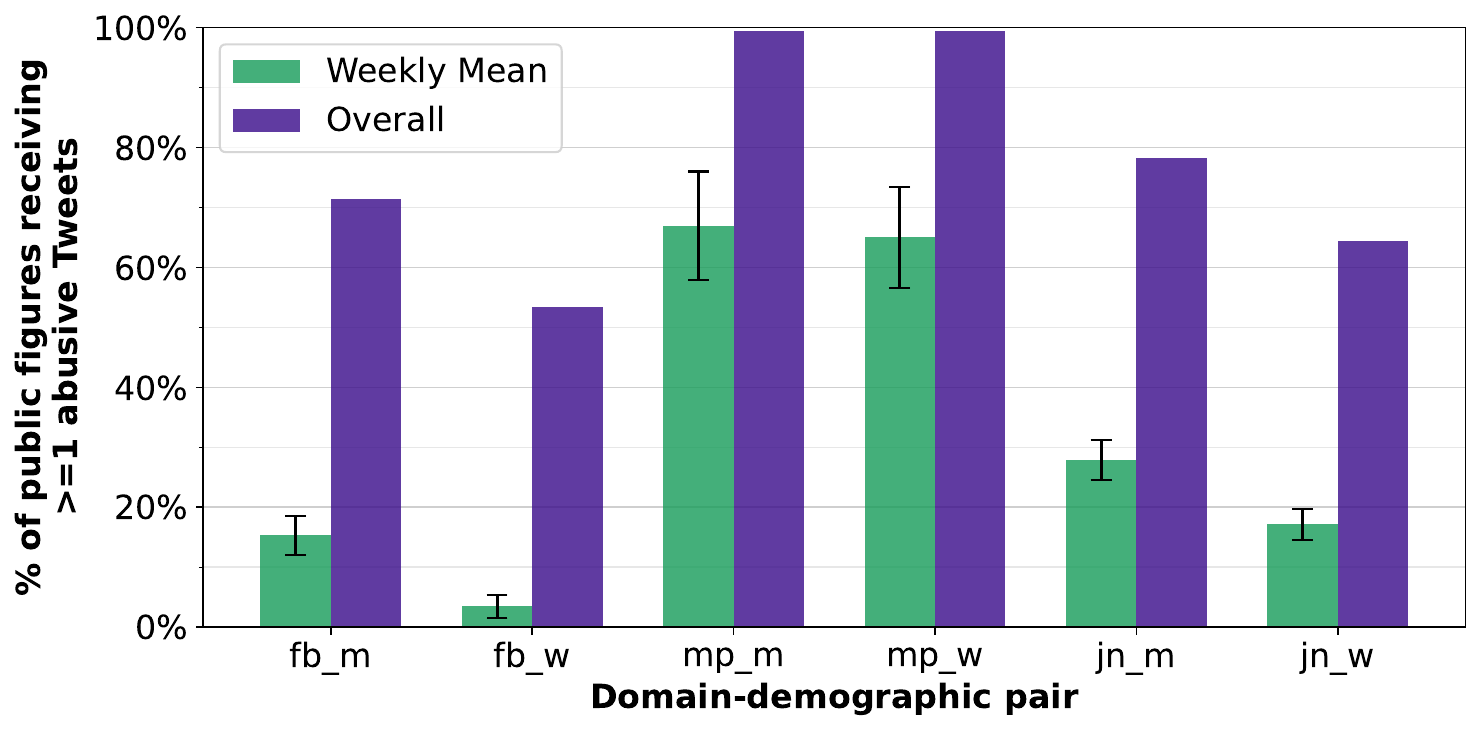}
    \caption{
    {Weekly average and overall proportions of public figures receiving at least one abusive tweet}. \\ \small Shows standard deviation, and overall proportion of public figures who received at least one abusive tweet during all data collection.} \label{fig:pct_receiving_bars}
    % \end{adjustwidth}
\end{figure}

The proportion of public figures that received any abuse across the entire data collection window is visible in \cref{fig:pct_receiving_bars}, showing that almost all MPs (99.5\% for men and women) received any abuse, and over 50\% of footballers and journalists. The different lengths of the data collection windows does affect this, meaning we might see higher levels of coverage for journalists if the data collection window was more similar to that of MPs and footballers, but the fact that relatively few footballers (71.5\% of men, 53.4\% of women, lowest across both demographics) received any abuse, despite the data collection window being the largest of the 3 domains, does emphasise that fewer footballers receive any abuse at all than the other domains.

One might assume that the most abused public figures are also the most popular public figures. We use the Spearman rank correlation between the quantity of abuse a public figure receives and the number of followers they have, visible in \cref{table:popularity}. 

We see a mild to strong rank correlation between abuse and followers when considering the total population of each group of public figures, highest for footballers (0.83 for men, 0.71 for women). However, when we limit the analysis to only the top 50 most abused public figures from each group, this changes - there is still a positive rank correlation for all groups, but the correlation is significantly weaker for footballers and male journalists. This indicates that some of the most abused individuals within these groups have lower follower counts. This points to more circumstantial abuse centered around specific events, and more work is needed to understand this phenomenon.

\begin{table}[!h]
\begin{adjustwidth}{0in}{0in}
\centering
\begin{talltblr}[caption={Spearman rank correlation coefficients between abuse received and number of followers.}, label={table:popularity}]{
  % width = \linewidth,
  % colspec = {Q[238]Q[302]Q[150]Q[227]},
  row{2} = {r},
  cell{1}{1} = {r=2}{},
  cell{1}{2} = {r=2}{},
  cell{1}{3} = {c=2}{},
  cell{3}{1} = {r=2}{},
  cell{3}{3} = {r},
  cell{3}{4} = {r},
  cell{4}{3} = {r},
  cell{4}{4} = {r},
  cell{5}{1} = {r=2}{},
  cell{5}{3} = {r},
  cell{5}{4} = {r},
  cell{6}{3} = {r},
  cell{6}{4} = {r},
  cell{7}{1} = {r=2}{},
  cell{7}{3} = {r},
  cell{7}{4} = {r},
  cell{8}{3} = {r},
  cell{8}{4} = {r},
  hline{1,3,9} = {-}{},
  hline{2} = {3-4}{Gray},
  hline{5,7} = {-}{Gray},
}
\textbf{Domain} & \textbf{Demographic} & \textbf{Rank Correlation} &                 \\
                &                      & \textbf{All}              & \textbf{Top 50} \\
Footballers     & Men                  & 0.83                      & 0.39            \\
                & Women                & 0.71                      & 0.52            \\
MPs             & Men                  & 0.53                      & 0.45            \\
                & Women                & 0.49                      & 0.60            \\
Journalists     & Men                  & 0.57                      & 0.23            \\
                & Women                & 0.48                      & 0.51            
\end{talltblr}
\begin{tablenotes}
  \noindent\centering Coefficients are presented for all public figures within \\a group, and for the top 50 in terms of abuse received.
\end{tablenotes}
\end{adjustwidth}
\end{table}

\subsection{Temporal patterns}
\label{section:temporal_patterns}

It is well known that abuse of public figures is not a stable phenomenon temporally \cite{vidgen_tracking_2022}, with abuse rising and falling in relation to real world events. Here, we explore the extent to which abuse levels fluctuate over time, and how the dynamics of temporal fluctuation differ by domain and demographic.

\begin{figure}[!h]
    \begin{adjustwidth}{-0.2in}{0in}
    \includegraphics[width=1.05\linewidth]{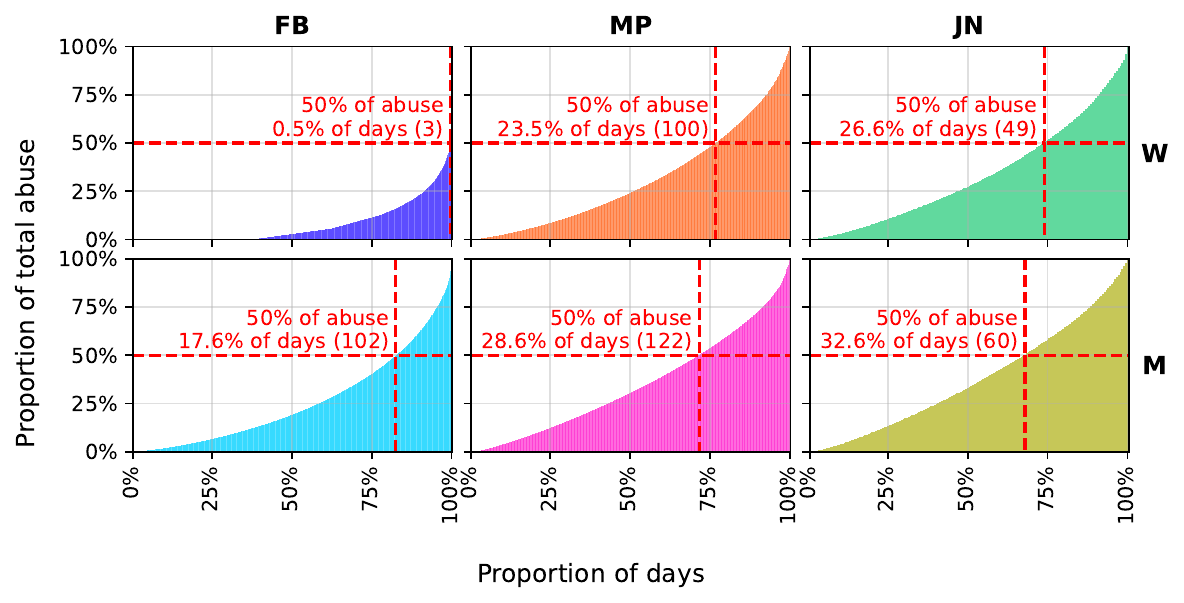}
    \caption{{Cumulative distributions of abuse by day.} \\ \small One column per domain, one row per gender. Proportion of the required top $N\%$ of days to reach 50\% of all abuse is annotated.}\label{fig:daydist}
    \end{adjustwidth}
\end{figure}

We present cumulative distributions of total abuse counts by day in \cref{fig:daydist}. We see that abuse tends to be unevenly distributed over time for all groups. 50\% of abuse targeted at footballers takes place over 9.1\% of days (on average across both men and women). Abuse towards MPs and journalists is more evenly distributed, with 50\% of abuse taking place over 26.1\% and 29.6\% of days respectively on average. Within each domain, abuse of female public figures tends to be more uneven than for their male counterparts. Female footballers receive 50\% of their abuse over just 0.5\% of days compared to 17.6\% of days for male footballers. This observation holds to a lesser extent for MPs and journalists, where the difference between female and male public figures is 5.1\% and 6.0\% respectively.

\begin{figure}[!h]
    \includegraphics[width=\linewidth]{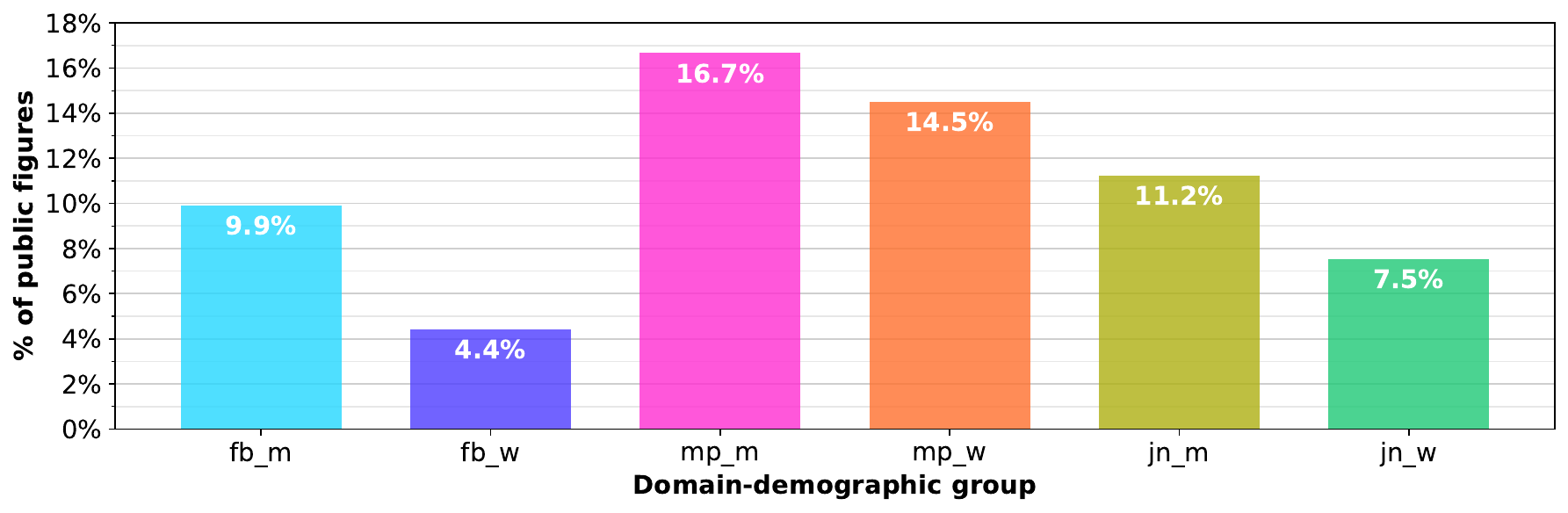}
    \caption{{Percentages of public figures who received $>=\frac{1}{3}$ of their total abuse in one day.} \\ \small Only includes public figures who received at least 10 abusive tweets during data collection.}\label{fig:x_y_in_z}
\end{figure}

Investigating temporal fluctuation at an individual level, we present percentages of public figures who receive at least $1\over3$ of their total abuse in a single day in \cref{fig:x_y_in_z}. This shows the presence of public figures in all groups who receive a significant proportion of the abuse they receive throughout the whole study in a single day. More MPs than any other group receive over $1\over3$ of their abuse in a single day (15.6\% on average), followed by journalists (9.4\% on average), and then footballers (7.2\% on average). Across all domains, more men receive a significant proportion of their abuse in a single day than women. 

\begin{figure}[!t]
    \includegraphics[width=\linewidth]{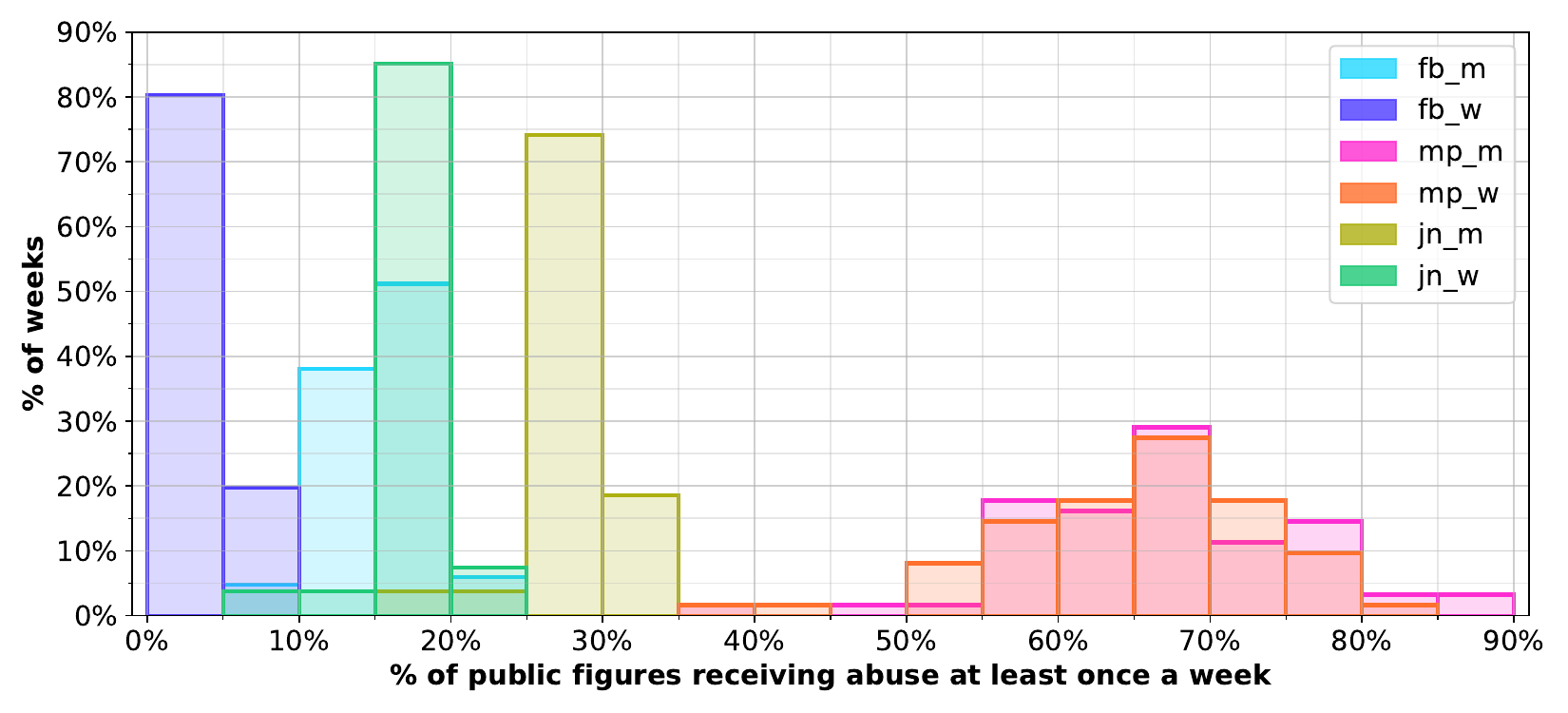}
    \caption{
    {Histograms of percentages of public figures receiving at least one abusive tweet in a week.}}
    \label{fig:pct_receiving}
\end{figure}

Finally, in \cref{fig:pct_receiving} we plot histograms of weekly percentages of public figures receiving any abuse within a given week. We see that the majority of MPs receive abuse on a weekly basis, with over 50\% of MPs receiving at least one abusive tweet in all but 2 (3.2\%) weeks during data collection. No other group receives such regular and widespread abuse - at no point do over 50\% of footballers or journalists in our dataset receive at least one abusive tweet in a single week,  with the average being 15.3\% for male footballers, 3.47\% for female footballers, 28.0\% for male journalists, and 17.2\% for female journalists, compared to 67.0\% for male MPs and 65.0\% for female MPs.

Taken together, we make several observations about these results. Firstly, MPs as a group receive the most regular and widespread abuse, but also individually see many concentrated periods of abuse. This suggests that whilst abuse may be a stable feature of being an MP, there is also a large element of this abuse which is more unstable and sporadic. These abusive tweets could be in relation to specific events, for example, in response to a controversial tweet or comment made by an MP. 

Footballers, on the other hand, receive the least regular and widespread abuse as a group. In most weeks a small proportion of players receive any abuse, which tends to be less evenly distributed over time. This suggests that abuse towards footballers is more sporadic and event driven. However, compared to MPs, a smaller minority of players receive a significant proportion of their abuse in a single day. It may be the case that whilst footballers receive less regular abuse than MPs \textit{as a group}, many individual footballers receive more regular abuse spread out over specific days, such as match days. One can imagine this as a series of regular peaks in abuse between troughs of low abuse levels, resulting in an uneven distribution but lacking individual peaks that account for significant proportions of abuse.

The temporal nature of abuse towards journalists is somewhere between that of MPs and footballers. In most weeks a reasonable proportion of journalists receive some abuse, which is more evenly distributed over time than for MPs or footballers. The proportion of journalists who receive at least a third of their abuse in a single day (9.35\% on average across men and women) is less than MPs (15.6\% on average) but greater than footballers (7.15\% on average). 

On an individual level, less women receive over $1\over3$ of their abuse in a single day than men (across all domains), suggesting a more even distribution for abuse of female public figures. However, on a group level, abuse towards women is in fact less evenly distributed over time than for men. This requires more analysis to understand, but may be due to the presence of events that see female public figures abused as a group, to a greater extent than for male public figures.

\subsection{Factors linked to abuse}

We tackle our third and final research question regarding which factors are linked to abuse. We firstly examine the relationship between the activity of public figures and the abuse they receive, and then attempt to quantify how intrinsic abuse is to a domain or gender through statistical modelling.

\begin{figure*}[!t]
    \begin{adjustwidth}{-0.5in}{0in}
    \includegraphics[width=1.1\linewidth]{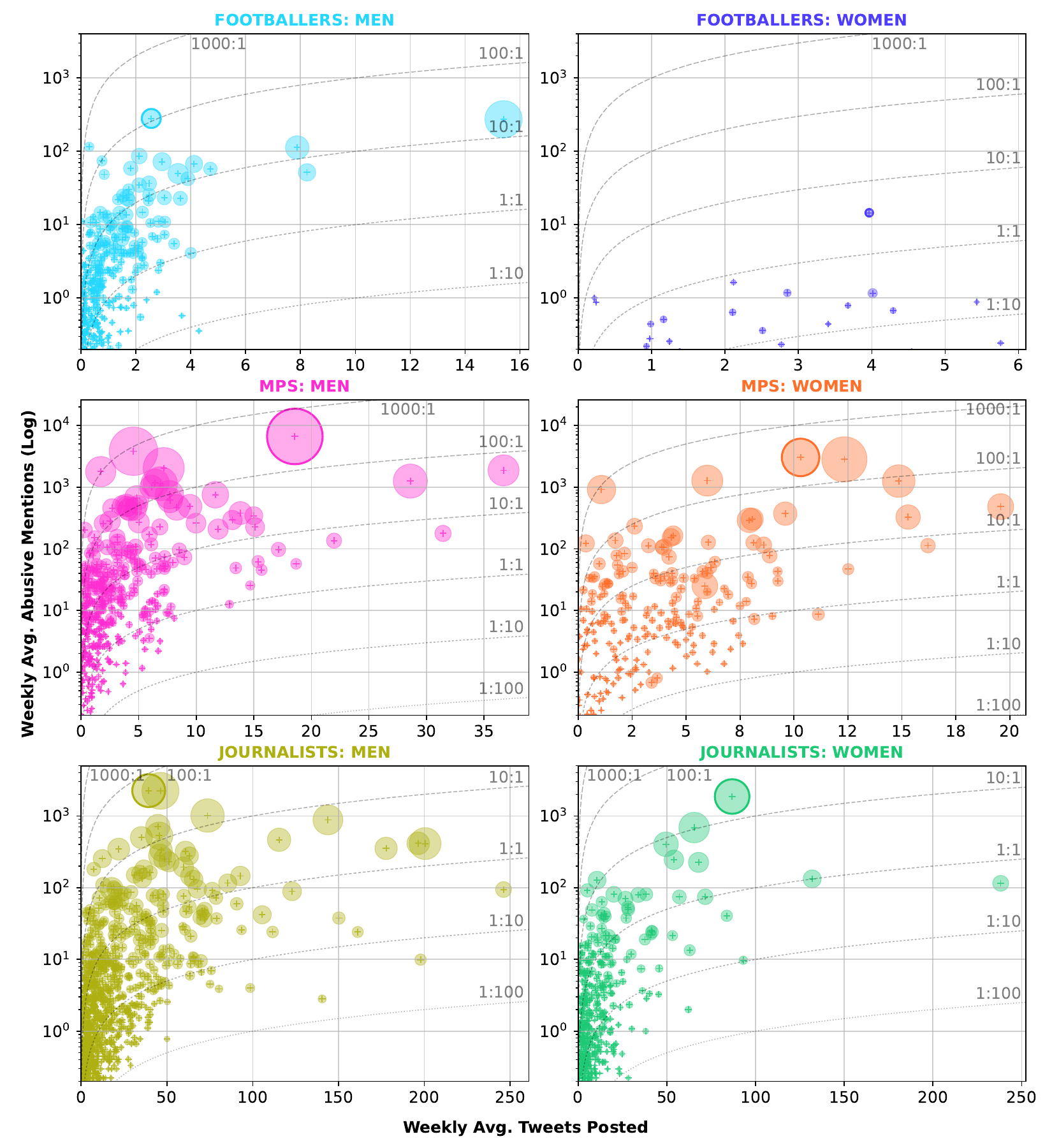}
    \caption{{Public figure activity against abuse received.} \\ \small Dotted lines represent ratios of abusive mentions to Tweets posted by a public figure. Size of markers represents total tweets received by the public figure. The most abused public figure in each domain is marked with a solid outline.}
    \label{fig:activity}
    \end{adjustwidth}
\end{figure*}

\subsubsection{Public figure activity}

We count the number of ``active statuses'' written by a public figure (the number of statuses posted by a public figure account that receive at least one reply) as a measure of their activity. 

Looking at weekly average activity in \cref{table:wkavg_metrics}, journalists appear as the most active group by a significant factor, with MPs more active than footballers on average by a smaller margin. This holds when accounting for the number of public figures studied, given the larger number of journalists - the average journalist writes 7.2 (9.7 for men, 4.6 for women) tweets per week, compared to 3.3 (3.1 for men, 3.6 for women) for the average MP and 0.6 (0.8 for men, 0.5 for women) for the average footballer. 
 
\cref{fig:activity} visualises the relationship between activity and abuse for public figures across different domains. In terms of ratios of abuse to activity, only considering public figures receiving at least 1 abusive tweet per week on average, we see 28 male and 5 female MPs receiving at least 100 abusive tweets for every tweet they write per week, compared to 9 and 4 journalists, and 3 and 0 footballers. The average ratio is also higher for MPs, at 32.1 abusive tweets per status for men and 22.7 for women, versus 17.2 and 1.9 for footballers and 1.4 and 1.6 for journalists. 

There appears to be some positive relationship between activity and abuse. Correlation coefficients between abuse and activity are mild but positive, with the average Pearson correlation coefficients at 0.32, strongest for male footballers (0.64). However, across all groups, the most active individual is never the most abused individual, but does still consistently rank in the top 8\% of public figures. This suggests that, while some level of activity may be a pre-requisite to receiving higher levels of abuse, it doesn't necessarily identify the most abused individuals.

\subsubsection{Intrinsic nature of abuse}

In \cref{section:dist_of_abuse} and \cref{section:temporal_patterns} we use \textit{absolute} levels of abuse to compare across domains, demographics, and time. This is important as it affords us a better understanding of how public figures actually experience abuse online. However, this approach does not allow us to assess the extent to which abuse is \textit{intrinsic} to a domain or demographic -- that is, whether abuse is a direct result of belonging to a particular domain or demographic, or whether it can be explained by other factors irrespective of domain or demographic (such as the prominence or activity of a public figure).

To examine the intrinsic nature of abuse we fit a series of count models. Each observation in these models is a public figure and the dependent variable is the number of abusive tweets received by that individual. We first fit models to each domain separately to examine gender differences within each domain (referred to as Model 1 for Journalists, Model 2 for MPs, and Model 3 for Footballers). In these models, our main independent variable of interest is \textit{gender}, where female gender is used as the reference category. We subsequently fit a model to all the data to assess differences between domains (referred to as Model 4). Here, we are interested in the independent variable \textit{domain}, where the footballer domain is used as the reference category. We also include in Model 4 an exposure offset term to account for the differing time periods in which data were collected for each domain. This was set to the log of the total number of weeks of data collection for each domain. In all models we include as control variables the total number of audience contact tweets received by the public figure (\textit{Count total}), the number of people that follow them (\textit{Count followers}), and the number of tweets written by the public figure that received at least one reply (\textit{Count replied to}). To assist with model convergence, and to aid interpretation of results, these variables were incremented by 1 and log\textsubscript{2} transformed.

For each model, we use a likelihood ratio test (LRT) to determine whether a Poisson or a negative binomial regression is most appropriate. As the Poisson model is nested within the negative binomial model, the LRT is a suitable test to compare the fit of these models. All tests indicated the negative binomial provided a significantly better fit. As the negative binomial model estimates a dispersion parameter, which is held constant in the Poisson model, this suggests our data is over-dispersed. We considered using zero-inflated versions of these models (i.e. zero-inflated Poisson and zero-inflated negative binomial), but the lack of a strong theoretical reason for the existence of excess zeros deemed these inappropriate. Negative binomial models were run using the \texttt{MASS} R package \cite{mass}, and approximate 95\% confidence intervals were obtained by likelihood profiling. We report incident rate ratios (IRRs) by exponentiating the raw model coefficients. For a log\textsubscript{2} transformed count variable, the resulting IRR represents the multiplicative change in incident rate when that count is doubled.

Models were checked for multicolinearity by calculating variance inflation factors (VIFs), with VIFs for all variables in all models no greater than 5. Outliers were also checked for using Cook’s distance (CD). A cutoff of $4 / N$ (where $N$ is the number of observations) is typically used to identify potential outliers. Data points with a CD greater than or equal to this cutoff were removed, models refit, and the resulting coefficients checked for changes in direction and significance. Two changes were observed after refitting models. In Model 3, the control variable \textit{Count replied to} was no longer deemed significant, whilst in Model 4 the raw estimate for \textit{Count followers} changed from -0.002 to 0.003 (representing a change in the corresponding IRR from 0.998 to 1.003)\footnote{We note that 216 potential outliers were observed for Model 4. After carefully inspecting these data points and concluding they were genuine (and not as a result of data errors) we decided against excluding them from our analysis.}.Taken together, these diagnostics suggest no reason to doubt the reported results. We present results for all models in Table \ref{table:regression_results}.

\begin{table}[!h]
\setlength\extrarowheight{2.5pt}
    \caption{Negative binomial model results.}\label{table:regression_results}
    \begin{tabular}{lcc}
    \hline
    \textbf{Term}            & \textbf{IRR}         & \textbf{95\% Conf. Interval} \\ \hline
    \textbf{Model 1 (Journalists)} &                      &                                   \\
    Count total              & 2.721                & 2.643 - 2.803                     \\
    Count followers          & 0.936                & 0.911 - 0.962                     \\
    Count replied to         & 0.860                & 0.835 - 0.885                     \\
    Gender (male)                 & 1.219                & 1.135 - 1.310                     \\
                             & \multicolumn{1}{l}{} & \multicolumn{1}{l}{}              \\
    \textbf{Model 2 (MPs)} &                      &                                   \\
    Count total              & 2.576               & 2.497 - 2.658                     \\
    Count followers          & 0.893                & 0.863 - 0.924                     \\
    Count replied to         & 0.942                & 0.912 - 0.972                     \\
    Gender (male)                 & 1.259                & 1.146 - 1.383                     \\
                             & \multicolumn{1}{l}{} & \multicolumn{1}{l}{}              \\
    \textbf{Model 3 (Footballers)} &                      &                                   \\
    Count total              & 2.201                & 2.098 - 2.312                     \\
    Count followers          & 1.094                & 1.053 - 1.137                     \\
    Count replied to         & 0.907                & 0.857 - 0.959                     \\
    Gender (male)                 & 2.792                & 2.209 - 3.521                     \\
                             & \multicolumn{1}{l}{} & \multicolumn{1}{l}{}              \\
    \textbf{Model 4 (All)} & \multicolumn{1}{l}{} & \multicolumn{1}{l}{}              \\
    Count total              & 2.555                & 2.502 - 2.610                     \\
    Count followers          & 0.998                & 0.980 - 1.017                     \\
    Count replied to         & 0.877                & 0.858 - 0.896                     \\
    Gender male                 & 1.388                & 1.307 - 1.473                     \\
    Domain (journalists)               & 11.721               & 10.638 - 12.911                   \\
    Domain (MPs)               & 5.492                & 4.891 - 6.168                     \\ \hline
    \end{tabular}
\end{table}

Models 1 – 3 look at how the levels of abuse received by public figures varies by gender. In all three models male public figures experience more abuse than female public figures. On average, and with all other variables held constant, male journalists receive a 22\% greater incidence of abusive tweets than their female counterparts, whilst male MPs receive a 26\% greater incidence. Male footballers receive an incidence that is almost three times greater than for female footballers. The 95\% confidence intervals do not contain 1, and so these estimates can be considered statistically significant. 

Model 4 looks at how abuse levels differ between domains. The results show that on average, and with all other variables held constant, journalists receive abuse at a rate which is almost 12 times that for footballers. The rate at which MPs receive abuse is almost 6 times that for footballers. Again, since the confidence intervals do not contain 1, these estimates can be considered statistically significant at the 95\% level.

\section{Discussion and conclusion}\label{section:discussion}
Overall, we find that MPs receive higher absolute levels of abuse at a more constant rate than footballers or journalists, although abuse appears to be a greater intrinsic feature of being a journalist than it is for an MP or footballer. Across all domains, the majority of abuse is directed at a very small number of individuals, but the majority of all public figures studied across each group received at 1 least abusive tweet during the data collection window. Abuse levels are more evenly distributed over time, but fluctuate to a much greater degree for footballers than MPs or journalists. We find that abuse levels tend to be higher for public figures who are more active or have more followers, but the most abused individuals are rarely the most active or most followed. Across all domains, an average male public figure receives more abuse than an average female public figure, but also has more followers and receives more tweets in total - controlling for these factors, statistical models still indicate that being a man is predictive of higher abuse levels. We also see that abuse targeted at female public figures fluctuates with time to a greater extent than male public figures, and is therefore likely to be driven by specific events.

\subsection{Limitations}
\label{section:limitations}

In this study we focus on a broad-reaching definition of abuse, and count the number of tweets that meet that definition (as classified by a machine learning model). This does not account for the potential range of severity of abuse, and as such all abuse counts equally towards the figures presented, lacking the nuance that some public figures may receive higher levels of more severe abuse than others. In the same vein, abuse may affect individuals in different ways, and measuring counts of tweets does not encapsulate the impact of abuse on individuals. As such, our results are best interpreted as counts of abusive language, with further work needed to understand how the severity and impact of abuse differs across groups of public figures.

Our focus on data collection of public figures enables relatively efficient data collection, and our filtering of tweets to the ``audience contact'' category maximises the chance of any given tweet being directly addressed at a public figure. However, we do identify cases of e.g. abusive replies to public figures that in fact show support . Equally, abuse doesn't solely exist within this category on social media - many public figures receive abuse via direct messages, which are not accessible to , and abuse may also take place without mention or reply to the subject of the abuse. As noted earlier, this study is limited to a single platform, and as such conclusions can only be drawn within the scope of Twitter/X.

We delineate public figures into binary gender categories. As noted, we do not include other possible gender identities due to low prevalence within the groups studied. A public figure from a minority gender identity is likely to receive abuse targeted surrounding their identity, and abuse targeted at these individuals is likely to follow different dynamics.

This study was conducted sequentially, with data collection, annotation, and modelling occurring at different times for different domain-demographic groups. Data collection via archive search may not include tweets that would have been obtained during streaming. Data annotation uses the same schema across all groups, but, as discussed, annotation is done by 2 different groups (the expert annotator group remained the same), which may introduce uncertainty. The first model trained in this study utilised active learning, an effective but resource-intensive approach which could not be replicated for later models.

The variable time windows in data collection in this study represent variation between the groups studied. Arguably the real world events that occur within these time windows skew results, but one would struggle to be able to measure a ``baseline'' level of abuse through real world data collection. We take measures to account for variable times windows at multiple points in the analysis.

Our annotation schema (\cref{fig:annotinstructions}) differentiates between criticism and abuse. Much of the abuse received by public figures could be seen as overly-profane or toxic forms of criticism, highlighting the fuzzy line between the two categories. While annotators were given a strict schema to follow, we note that levels of annotator disagreement were higher in cases where the final majority label was either ``critical'' or ``abusive''.

\subsection{Future Work}
\label{section:futurework}

As discussed, expanding beyond binary gender the would be a logical extension to this work. In addition to this, further work could be done to expand beyond a single demographic (gender) to better understand the dynamics of abuse across a range of identities.

Content analysis (with more nuance than whether content is abusive or not according to a machine learning model) of tweets targeted at public figures would provide a greater understanding of the themes contained within abuse, and could be combined with incorporation of data around real world events to provide more granular explanations of specific peaks in abuse.

Further research questions building on this work include developing a better understanding of the perpetrators of abuse, and how the affiliations and beliefs of perpetrators of abuse varies between groups of public figures. Extending beyond binary abuse classification to be able to measure the severity of abuse, and discern different forms of abuse (e.g. misogyny, racism), would open up avenues to explore the types of abuse received by public figures from different domains and demographics.

\newpage

%%%% BACKMATTER
\backmatter

\section*{Declarations}

\subsection*{Availability of data and materials}
The dataset of tweets analysed during the current study is not publicly available due to restrictions on sharing data collected from the Twitter API, as outlined in the API terms and conditions. We make available anonymised aggregate statistics, available from the corresponding author on reasonable request.

\subsection*{Competing interests}
The authors declare that they have no competing interests.

\subsection*{Funding}
This work was partially supported by the Ecosystem Leadership Award under the EPSRC Grant EPX03870X1 and The Alan Turing Institute.

\subsection*{Authors' contributions}
LBM, AW, and JB designed the study. LBM and AW performed data collection and analysed the results. LBM, AW, and JB wrote the first draft. All authors read and approved the final manuscript.

\subsection*{Acknowledgements}
We thank Eirini Koutsouroupa for invaluable project management support, and Yi-Ling Chung, Ivan Debono, Pica Johansson, Hannah Kirk, Francesca Stevens, and Bertie Vidgen for their roles in previous work that enabled this study.

\newpage

\bibliography{sn-bibliography}

\newpage

\section*{Appendices}
\appendix
\section{Data collection} \label{appendix:datacollection}

\subsection{Sourcing public figures}
For footballers, we focus on the top UK leagues of the men's and women's game, namely the Men's Premier League, consisting of 20 teams at any given point, and the Women's Super League, with 12 teams at any given point. The exact number of players in each league is fuzzy, but we estimate the total number of eligible individuals to be around 1,000 male footballers and 300 female footballers. Within politics, we focus on UK Members of Parliament (MPs), the most prominent public figures in UK politics as elected officials voted for by the public, constituting 650 individuals.  Unlike footballers and MPs, journalists are not a finite group of individuals, adding complexity to the selection of individuals and data collection. As such, we use a list of UK journalists on Twitter \cite{jn_twitter_list} (now no longer maintained), selecting the top 3,000 journalists in terms of Twitter follower numbers. We chose a larger number of journalists than MPs or footballers to account for the fact that there is no finite number, and to capture a range of abuse, given that abuse is not a phenomenon limited to the most popular journalists.

\subsection{Collecting public figure information} We scrape the official websites of the Premier League \cite{fbm_list} and Super League \cite{fbw_list}, alongside the websites of individual clubs, to gather lists of eligible players. This information includes complete information on name, club, and nationality, and incomplete information on position, date of birth, height, and Twitter account (gender is implicit in the data collection process). For MPs, we collate a complete list using several sources \cite{mp_list1} \cite{mp_list2} \cite{mp_list3} (the latter is no longer available). These provide complete information on name, gender, party, constituency, and incomplete information on Twitter account. The list of journalists \cite{jn_twitter_list} provides complete information on name, publisher, publication, job role, and Twitter account, and no information on gender.

We estimate gender for journalists using a hybrid approach, first obtaining a gender and probability from the first names of the 3,000 journalists studied using genderize \cite{genderize} (based on census data). In cases where this approach returned a gender with a probability less than 100\% (912 entries, 30\%), we prompt GPT-4 \cite{openai_gpt4_2023} with the full name and publication of the journalist, asking to indicate if it is aware of the journalist and what their gender is. In cases where the two approaches disagreed or GPT-4 did not indicate awareness of the individual (313 entries, 10\%), we manually labelled gender using available resources online.

\subsection{Social media presence} Where records of Twitter profiles were incomplete (footballers and MPs), we use a combination of the Twitter API and desk research to assign Twitter profiles where they exist (some MPs and footballers are not present on Twitter). During data collection, changes in user name or deletions of account were recorded, as were any changes in affiliation (e.g. footballers transferring to other clubs, MPs losing their seat). Numbers of accounts and any analysis presented is inclusive of all accounts used for data collection through the entire data collection process.

\section{Annotator Instructions}

\begin{figure*}[!h]    
    \centering
    \begin{adjustwidth}{0in}{0in}
    \includegraphics[width=.81\textwidth]{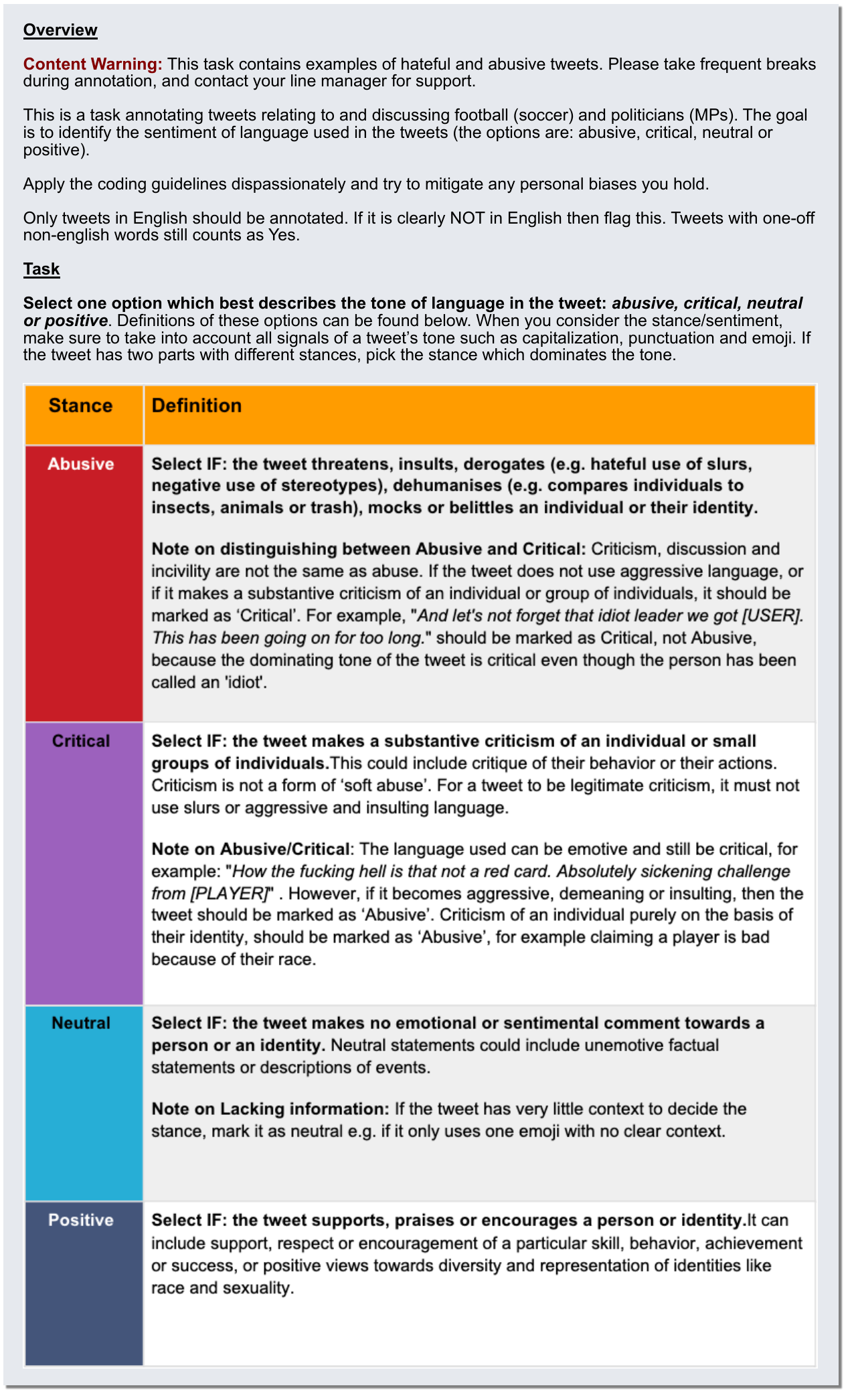}
    \caption{Instructions given to annotators.}
    \label{fig:annotinstructions}
    \end{adjustwidth}
\end{figure*}

\end{document}